\documentclass[conference]{IEEEtran}
\ifCLASSINFOpdf
   \usepackage[pdftex]{graphicx}
\else
   \usepackage[dvips]{graphicx}
\usepackage{hyperref}
\hypersetup{%
  colorlinks=true, 
  breaklinks=true, 
  urlcolor=blue, 
  linkcolor=blue, 
  citecolor=blue, 
} %%%

\fi
\usepackage[cmex10]{amsmath}
\begin{document}

\title{ \LARGE Repurposing Coal Power Plants into Thermal Energy Storage for Supporting Zero-carbon Data Centers}

%% To specify the authors when (number of affiliations <= 2)
\author{
\IEEEauthorblockN{Yifu Ding, Serena Patel, Dharik Mallapragada, Robert James Stoner}
\IEEEauthorblockA{MIT Energy Initiative, \{yifuding, serenap, dharik, stoner\}@mit.edu}
%\and
%\IEEEauthorblockN{Author n.1 Name per Affiliation B\\ Author n.2 Name per Affiliation B}
%\IEEEauthorblockA{(Affiliation B) Department Name of Organization \\
%Name of the organization, acronyms acceptable\\
%City, Country\\
%\{email author n.1, email author n.2\}@domain (if desired)}
}

%% To specify the authors when (number of affiliations > 2)
% \author{\IEEEauthorblockN{Author n.1\IEEEauthorrefmark{1},
% Author n.2\IEEEauthorrefmark{2},
% Author n.3\IEEEauthorrefmark{3}, 
% Author n.4\IEEEauthorrefmark{3} and
% Author n.5\IEEEauthorrefmark{4}}
% \IEEEauthorblockA{\IEEEauthorrefmark{1} Department Name of Organization A\\
% Name of the organization A,
% Address A\\ Emails if wanted}
% \IEEEauthorblockA{\IEEEauthorrefmark{2} Department Name of Organization B\\
% Name of the organization B,
% Address B\\ Emails if wanted}
% \IEEEauthorblockA{\IEEEauthorrefmark{3} Department Name of Organization C\\
% Name of the organization C,
% Address C\\ Emails if wanted}
% \IEEEauthorblockA{\IEEEauthorrefmark{4}Department Name of Organization D\\
% Name of the organization D,
% Address D\\ Emails if wanted}
% }

\maketitle
\thispagestyle{plain}
\pagestyle{plain}

\begin{abstract}

Coal power plants need to be phased out and face stranded asset risks under the net-zero energy system transition. Repurposing coal power plants could save costs and reduce carbon emissions using the existing infrastructure and grid connections. This paper investigates a retrofitting strategy that turns coal power plants into thermal energy storage (TES) and zero-carbon data centers (DCs). The proposed capacity expansion model considers the co-locations of DCs, local renewable generation, and energy storage with the system-level coal retirement and retrofitting. We optimize the DC system configurations under the hourly-matching carbon policy and flexible operations. Results show that under hourly-matching carbon constraints, the retrofitted TES could complement the operations of lithium-ion batteries (LIBs) to reduce system costs. This could render DCs with optimal co-located renewable generations and energy storage more cost-effective than unconstrained DCs.
 
\end{abstract}


\section{Introduction}

\subsection{Background and motivations}

Coal power plants are the largest carbon emission source in the world and contributed to 15\% of the total emissions in 2020. The total capacity of U.S. coal power plants reached 204 GW by 2020 \cite{eia_capacity_2023}. Decarbonizing power systems requires the early retirement of high-emitting coal power plants and the integration of renewable energy. However, coal power plants provide inertia and ancillary services in the power systems, and renewable capacity expansion requires significant investments and new grid interconnections. In fact, a number of renewable generation projects in the U.S. get delayed or even canceled when waiting for interconnection construction \cite{lawrence_berkeley_national_laboratory_electricity_market_and_policy_queued_2022}. 

Retrofitting coal power plants provides a cost-saving solution by reusing the existing infrastructure and interconnections. They can be repurposed into thermal energy storage (TES) \cite{yong_retrofitting_2022}, nuclear reactors \cite{hansen_investigating_2022}, and data centers (DCs) \cite{andrew_coffman_smith_plan_2023}. These projects could significantly reduce carbon footprint and facilitate renewable energy integration. For example, when retrofitting coal power plants into TES, the boiler is replaced by heat storage and heat exchangers to store energy. The power is discharged via power blocks such as steam turbines. The sensible heat storage can be built using low-cost storage materials such as molten salt and thus has a lower energy capital cost of \$15-25/kWh$_{th}$ \cite{sepulveda_design_2021}. This makes TES a potential technology for long-duration energy storage designs in renewable-heavy grids or off-grid systems \cite{albertus_long-duration_2020}.

Another promising approach to repurposing coal power plants is DCs. DCs are the backbone of IT services and data management. The total energy consumption of DCs worldwide had grown to 273 TWh by 2020, which accounted for 1\% of the global energy consumption \cite{masanet_recalibrating_2020}. Several projections show this percentage could increase to around 20-50\% by 2030 \cite{jones_how_2018, mytton_sources_2022}. The increasing share of hyperscale DCs, whose energy consumption is equivalent to thousands of households, has significantly stressed local electricity transmission and generation \cite{masanet_recalibrating_2020, mytton_sources_2022}. IT companies nowadays build DCs on the former sites of coal power plants to reuse interconnections and mitigate impacts, such as demonstration projects in New York State \cite {andrew_coffman_smith_plan_2023}.

%These projects have been built in New York City \cite{andrew_coffman_smith_plan_2023} and Wisconsin \cite{timothy_gardner_insight_2023}. 

\subsection{Related works}

To offset additional carbon emissions of DCs and meet the net-zero carbon goal, cloud providers voluntarily purchase green electricity to match their energy consumption via power purchase agreements and unbundled renewable energy certificates \cite{anna_cybulsky_producing_2023}. For example, Google has set the goal to run 24/7 carbon-free energy on every grid the company operates by 2030 \cite{google_data_center_247_nodate}. DCs also provide flexibility to utilize intermittent renewable energy and reduce curtailment. Previous works have proposed load-shifting and coordination strategies for DCs to reduce carbon emissions, including the spatial load migration \cite{zheng_mitigating_2020} and temporal computing workload shifting \cite{lin_adapting_2023}. Nevertheless, the prerequisite of their implementations is to plan adequate local renewable capacity and energy storage in tandem with the increasing load of DCs. 

DCs require high reliability in power supply. Commercial DCs only allow 2.4 minutes to 1.6 hours of downtime per year and have an uninterrupted power supply onsite for emergency \cite{ahmed_review_2021}. However, it is very cost-prohibitive to use only short-duration LIBs and renewable energy resources for a reliable power supply. Ref. \cite{gnibga_renewable_2023} discusses the optimal configurations of on- and off-grid DCs using renewable energy sources. It shows that a combination of hydrogen storage and LIBs will reduce costs for the off-grid DC systems, even though the total investment is still much higher than the grid-connected DCs.

To this end, retrofitting coal power plants into low-cost TES to support zero-carbon DCs brings a unique opportunity. This paper develops an integrated capacity expansion strategy that links plant-level retrofitting strategies with upstream grid flexibility needs. This model co-optimizes DCs, retrofitted TES, LIB, and local renewable generations and also investigates the system-wide economic and carbon emission impacts when there are flexible or inflexible DCs.

\section{Model formulations}

\subsection{Capacity expansion problem considering coal retrofitting}
We model a power system that has a main system and multiple regions $z \in \mathcal{Z}$ to co-locate DCs and generation capacity. It also includes a set of gas and coal power plants $y \in \mathcal{G}^{th}$, renewable energy resources $y \in \mathcal{G}^{res}$, and grid-connected energy storage $s \in \mathcal{O}$. The proposed capacity expansion problem considers investment, retirement, and retrofitting of coal power plants and investments in renewable resources. The model also sizes the grid-connected LIBs $ s \in \mathcal{O}^{lib}$ and retrofitted TES $ s \in \mathcal{O}^{tes}$. While the LIB has a symmetrical charging and discharging power capacity, TES is asymmetrical energy storage that requires the sizing of charge, discharge, and energy capacity. The objective function consists of capacity, retrofitting investments, and operation costs.

 \begin{equation} 
\begin{aligned}
	\text{min} \sum_{z \in \mathcal{Z}} \{&\underbrace{\sum_{y \in \mathcal{G}^{th}} I^{th}_{y}\overline{P}^{size}_{y}  \Omega^{inv}_{y, z} + F^{th}_{y} \overline{P}^{size}_{y}  \Omega^{net}_{y, z}}_{\text{Investment, retirement and retrofitting of coal power plants} } + \\\
 	&\underbrace{\sum_{y \in \mathcal{G}^{res}}I^{res}_{y} \overline{P}^{inv}_{y, z}  + F^{res}_{y} ( \overline{P}^{inv}_{y, z} + \overline{P}^{exist}_{y, z})}_{\text{Planning renewable resources} } +   \\\
 	&\underbrace{\sum_{s \in \mathcal{O}^{lib}}I^{dis}_{s} \overline{P}^{dis}_{s, z} + I^{en}_{s} \overline{E}^{inv}_{s, z} + F^{dis}_{s} ( \overline{P}^{inv}_{s, z} +  \overline{P}^{exist}_{s, z})}_{\text{Sizing LIBs} } + \\\
	&\underbrace{\sum_{s \in \mathcal{O}^{tes}} I^{ch}_{s} \overline{P}^{ch}_{s, z} + I^{retro}_{s} \overline{P}^{dis}_{s, z} + I^{en}_{s} \overline{E}^{tes}_{s, z} + F^{dis}_{s}  \overline{P}^{dis}_{s, z}}_{\text{Sizing retrofitted TES} } +  \\\
	& \underbrace{\sum_{y \in \mathcal{G}^{th}} \sum_{t \in \mathcal{T}}  \omega_{t}(\pi^{vom}_{y} + \pi^{fuel}_{y}) P^{dis}_{y,z,t}}_{\text{Varible O\&M costs of thermal power plants} } +  \\\
	&\underbrace{\sum_{s \in \mathcal{O}^{tes}}   \sum_{t \in \mathcal{T}} \omega_{t}\pi^{vom,dis}_{s} P^{dis}_{s, z, t}}_{\text{Variable O\&M costs of retrofitted TES} } + \underbrace{ \sum_{t \in \mathcal{T}}\omega_{t} \pi^{unmet}_{z, t} r^{unmet}_{z, t}}_{\text{Load shedding costs} }  \}
\end{aligned}
\label{tes_model}
\end{equation}

The power system operation is modeled by the hourly linear unit commitment at $t\in \mathcal T$. It has an integer variable $\Omega$ to represent thermal power plants' start-up and shut-down status, and the unit size is denoted as $\overline{P}^{size}$. We cluster the hourly renewable capacity factor data and load profiles into several representative weeks using the k-means clustering technique to solve the model swiftly. Then, we apply a weight $\omega_{t}$ on each hour of the representative weeks to calculate the full-year cost. The cycle costs of thermal power plants include variable operation and maintenance $ \pi^{vom}$ and fuel costs $ \pi^{fuel}$. The load shedding is based on the value of the lost load  $ \pi^{unmet}$.

The power outputs of thermal generations cannot exceed the net operating capacity and should be lower than the minimum stable generation level $\gamma_y$ .

 \begin{equation} 
\label{tes_model}
\gamma_y \overline{P}^{size}_{y} \Omega^{net}_{y, z} \leq P^{dis}_{y, z, t}  \leq \overline{P}^{size}_{y} \Omega^{net}_{y, z}  \quad y \in \mathcal{G}^{th}
\end{equation}

The net operating capacity of thermal generations considers the existing, newly built, retired, and retrofitted thermal generation units and their unit size. Coal power plants can only be retired or retrofitted, while natural gas power plants can be newly built or retired.

 \begin{equation} 
\label{tes_model}
\Omega^{net}_{y, z} := \Omega^{exist}_{y, z}  -  \Omega^{retire}_{y, z}  - \Omega^{retro}_{y, z}  \quad y \in \mathcal{G}^{coal}
\end{equation}

 \begin{equation} 
\label{tes_model}
\Omega^{net}_{y, z} := \Omega^{exist}_{y, z} +  \Omega^{inv}_{y, z}  -  \Omega^{retire}_{y, z}  \quad y \in \mathcal{G}^{gas} 
\end{equation}

This capacity expansion problem also incorporates ramp-up and-down constraints of thermal power plants and whole-system power balance, power, and energy balance of LIBs. These constraints are detailed in the GenX \cite{noauthor_genx_nodate}.

\subsection{Retrofitting coal power plants into TES}

We consider retrofitting coal power plants into the molten-salt TES, widely used in concentrating solar thermal power stations \cite{he_optimal_2016}. Its energy conversion processes involve three steps. First, TES is charged by the electrical heater using renewable energy (i.e., electricity-to-heat conversion). Then, the thermal energy is absorbed and stored in the molten salt tank for electricity generation at later periods. Finally, TES discharges power (i.e., heat-to-electricity conversion) via the steam turbine of the former coal plant known as the power block. The discharging capacity of the power block is no greater than the capacity of the former coal power plant.

 \begin{equation} 
\label{tes_model}
0 \leq \overline{P}^{dis}_{s,z} \leq \overline{P}^{size}_y  \Omega^{retro}_{y, z}  \quad s \in \mathcal{O}^{tes},  y \in \mathcal{G}^{coal}
\end{equation}

TES has an independent energy and power capacity, allowing simultaneous charging and discharging. Its charging efficiency $\eta^{ch}$ is determined by the resistive heat loss of the electric heater, and its discharging efficiency $\eta^{dis}$ is determined by the energy conversion rate of the steam turbine. The self-discharging rate $ \eta^{self}$ is applied due to the isolation of the thermal tanks. Assuming the initial state-of-charge (SoC) $E^{tes}_{s, 0}$, the energy balance of TES is given by,

 \begin{equation} 
\label{tes_model}
\begin{aligned}
E^{tes}_{s, t}  = E^{tes}_{s, t-1} + P^{ch}_{s, t}{\eta^{ch}}  {\Delta t}  - \frac{P^{dis}_{s, t}}{ \eta^{dis}} {\Delta t} -  \eta^{self} \overline{E}^{tes}_{s} \\\
 \quad s \in \mathcal{O}^{tes}, t \in [1, T]
\end{aligned}
\end{equation}

 We assume the SoC will not change with the ambient temperature and can be fully charged or discharged. 

 \begin{equation} 
\label{tes_model}
0 \leq E^{tes}_{s, t}  \leq \overline{E}^{tes}_{s}  \quad s \in \mathcal{O}^{tes}
\end{equation}

We define the maximum number of hours for discharging at the rated power as the battery duration between the maximum $d^{max}$ and the minimum durations $d^{min}$. 

  \begin{equation} 
\label{tes_duration}
h_s := \frac{ \overline{E}^{tes}_{s}} {\overline{P}^{dis}_{s}}  \quad s \in \mathcal{O}^{tes},  h_s \in [d^{min}, d^{max}]
\end{equation}

The power rate of the electric heater and steam turbine constrains the charging and discharging power of TES. We assume the minimum stable generation limit is relaxed after it is converted into TES.

  \begin{equation} 
\label{tes_model}
 0 \leq P^{ch}_{s, t}  \leq  \overline{P}^{ch}_{s} \quad s \in \mathcal{O}^{tes}, t \in \mathcal{T}
\end{equation}

  \begin{equation} 
\label{tes_model}
 0 \leq P^{dis}_{s, t}   \leq \overline{P}^{dis}_s \quad s \in \mathcal{O}^{tes}, t \in \mathcal{T}
\end{equation}

The power block of TES is modeled as a steam turbine, where the ramp-up and ramp-down rates, $\Delta P^{up}_y,  \Delta P^{down}_y $ are applied. 

 \begin{equation} 
\label{ramp_up}
  P^{dis}_{s, t} -   P^{dis}_{s, t-1} \leq \Delta P^{up}_y \overline{P}^{size}_y \quad s \in \mathcal{O}^{tes}, y \in \mathcal{G}^{coal}
, t \in [1, T]
\end{equation}

 \begin{equation} 
\label{ramp_up}
 P^{dis}_{s, t-1}  -   P^{dis}_{s, t}  \leq \Delta P^{down}_y \overline{P}^{size}_y  \quad s \in \mathcal{O}^{tes}, y \in \mathcal{G}^{coal}
, t \in [1, T]
\end{equation}

The fixed O\&M costs of TES are assumed to be the same as retrofitted coal power plants.

\subsection{Zero-carbon and flexible operations of DCs} In each co-location region  $z \in \mathcal{Z}$, zero-carbon DCs require the procured renewable electricity and energy storage assets to match their hourly power consumption. In addition, considering DCs could have temporal flexibility, a fraction of the DC workload can generally be deferred or shifted daily. 

Given a fraction $ \lambda_{z}  \in [0, 1]$ of inflexible computing loads, the hourly time-matching requirement constraint enforces that for every hour of the year, the DC power consumption is no greater than generations from local renewable capacity and the net injection from eligible energy storage. The hourly renewable generation is modeled by the renewable capacity factor $\alpha^{res}_{z, t}$ and additional local renewable capacity $\overline{P}^{inv}_{z, y} $. 

 \begin{equation} 
\label{hourly_matching}
\begin{aligned}
\sum_{y\in\mathcal{G}^{res}}   \alpha^{res}_{z, t} \overline{P}^{inv}_{z, y}  +   \sum_{s\in\mathcal{O}}  (P^{dis}_{z, s, t}   -   P^{ch}_{z, s, t}) \geq \lambda_{z} D^{dc}_{z, t}  \\\ \quad  t \in \mathcal T
\end{aligned}
\end{equation}

At each time step, the power to charge energy storage cannot exceed the maximum available local renewable generation.

 \begin{equation} 
\label{charging_control}
 \sum_{s\in\mathcal{O}^{tes}} P^{ch}_{z, s, t}  \leq \sum_{y\in\mathcal{G}^{res}} \alpha^{res}_{z, t} \overline{P}^{inv}_{z, y}   \quad t \in \mathcal T
\end{equation}

During one day, $t\in[t_0, t_0+H]$, the procured local renewable energy generations with energy storage must be greater than the daily energy consumption of DCs considering the temporal flexibility.

 \begin{equation} 
\label{flexibility}
\begin{aligned}
 \sum_{t\in[t_0, t_0+H]} (\sum_{y\in\mathcal{G}^{res}}   \alpha^{res}_{z, t} \overline{P}^{inv}_{z, y} +   \sum_{s\in\mathcal{O}} ( P^{dis}_{z, s, t}  -  P^{ch}_{z, s, t}))  \\\ \geq  \sum_{t\in[t_0, t_0+H]} D^{dc}_{z, t} 
\end{aligned}
\end{equation}

\section{Case study}

 A case study is conducted based on the ERCOT system. Ref. \cite{anna_cybulsky_producing_2023} details its costs and technical parameters. Hourly local renewable capacity factors are taken from Renewable ninja \cite{staffell_using_2016} around coal plants. 

\begin{figure}[!h]
    \centering
    \includegraphics[width=3.5in]{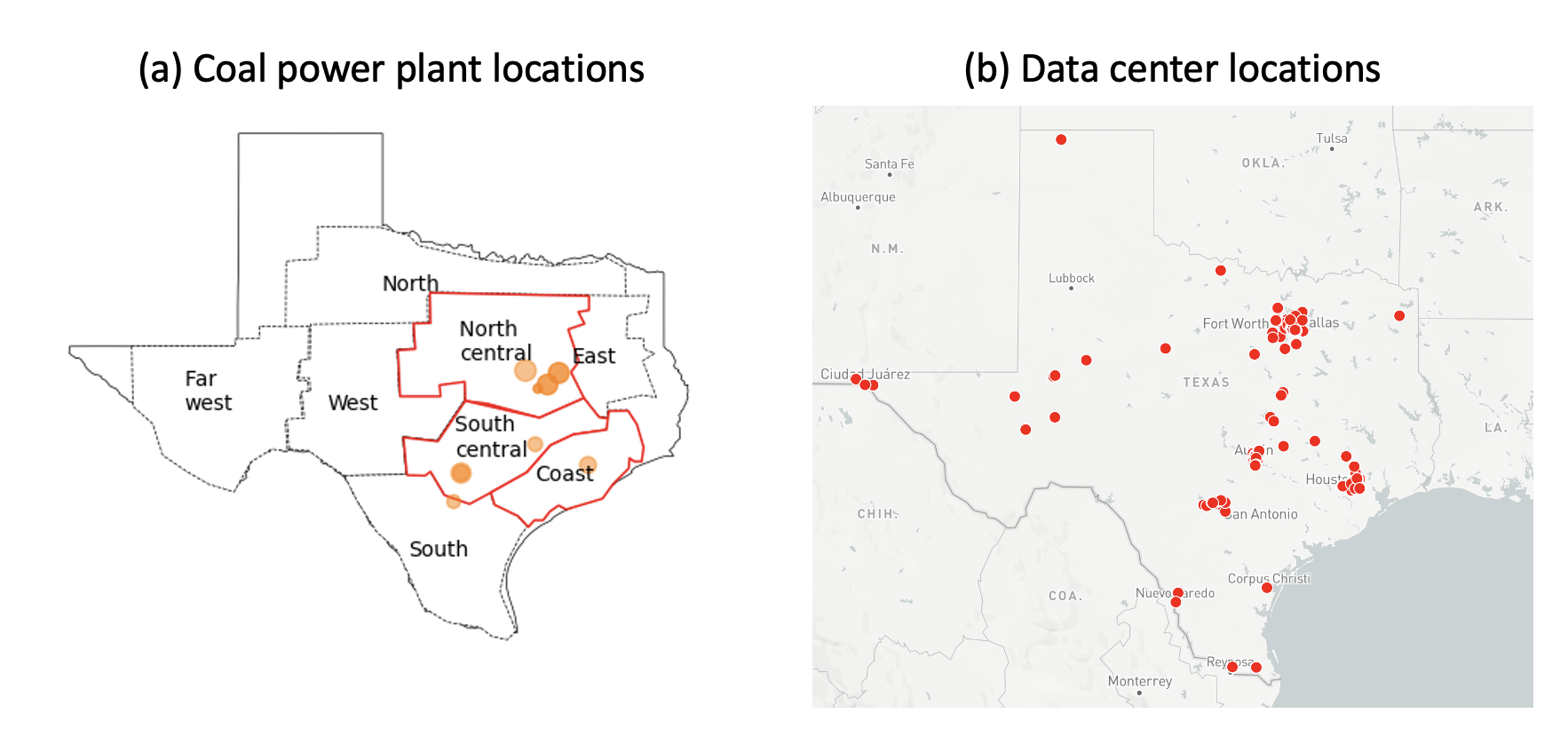}
        \caption{The locations of (a) operating coal plants of three weather zones highlighted in red \cite{us_energy_information_administration_coal-fired_nodate} and (b) the existing DCs \cite{baxtel_texas_nodate} as of 2022 in Texas}
    \label{retrofitted_coal}
\end{figure}

1) \textit{Coal power plants and DC loads:} We consider that the operating coal power plants in the ERCOT system as of 2022, except for those scheduled to be retired by 2025, are eligible for retrofitting. The sites of 12 eligible coal power plants (Fig. \ref{retrofitted_coal} (a)) highly overlap Texas's existing DC sites (Fig. \ref{retrofitted_coal} (b)). We collect their station heat values, remaining lifetime, and power capacity from the U.S. EIA \cite{us_energy_information_administration_coal-fired_nodate}. 

Texas has increasing DC demands in major cities. Based on the projections in ref. \cite{joshua_d_rhodes_impacts_2021}, we assume that three 1GW DCs will be built by 2030 in the North central, South central, and Coast zones of Texas (Fig. \ref{retrofitted_coal} (a)). DC loads have little seasonal or weather variation and thus are approximated as a flat load.

\begin{figure}[!h]
    \centering
    \includegraphics[width=3in]{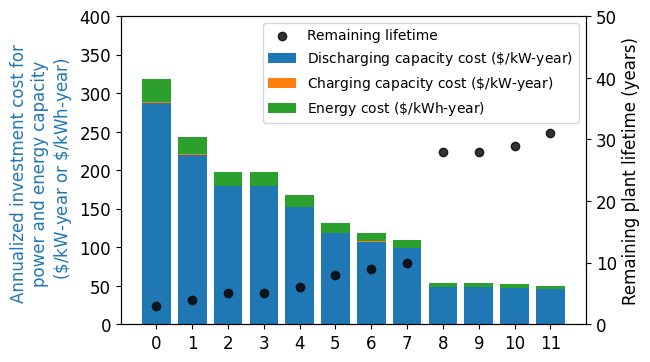}
        \caption{The annualized investment costs for charging, discharging, and energy capacity of the retrofitted TES considering the remaining lifetime of 12 eligible coal power plants by 2030}
    \label{retrofitted_coal_investment_cost}
\end{figure}

\begin{table}[!h]
\caption{Retrofitted TES parameters \cite{eia_capital_2020}}
\label{TES_cost_parameters}
\centering
\begin{tabular}{lccc} \hline
&Energy & Charge & Discharge \\  \hline
Technology &Molten-salt tank &Electric heater & Turbine\\
Investment & $\$$28.7/kWh$_{th}$ &$\$2$/kW &$\$799$/kW\\
Efficiency &$0.1\%$ &$35\% $& $95 \%$  \\ \hline
\end{tabular}
\end{table}

2) \textit{Cost assumptions of retrofitted TES and LIB:} Table \ref{TES_cost_parameters} presents the retrofitting technology, investment, and efficiency for energy conversion processes of the molten-salt TES. Assuming a 50-year lifetime of U.S. coal power plants, the remaining life of 12 coal power plants ranges from 2 to 33 years by 2030. Fig. \ref{retrofitted_coal_investment_cost} shows annualized investment costs of retrofitted TES calculated from capacity recovery factor at an annual interest rate $r = 4\%$. We assume the cost for component replacement is the half of new installation ($\$$1597/kW \cite{eia_capital_2020}). We set the duration of TES between 4 to 100 hours and the duration of LIBs to be four hours.

\section{Results}

The simulated scenarios include (1) different energy storage (i.e., LIB only, TES only, LIB \& TES) and (2) zero-carbon DCs that match their energy consumption with procured renewable energy v.s. Unconstrained DCs. (3) DC with v.s. without temporal flexibility.  The summary of all scenario results and inputs is in \cite{ding_supplement_nodate}.

%\begin{figure}[!h]
%    \centering
%    \includegraphics[width=3.5in]{image/changes_in_energy_barchart.png}
%        \caption{Changes in electricity generations (GWh) in six scenarios compared to the baseline}
%    \label{generation_compare}
%\end{figure}

1) \textit{Zero-carbon (ZC) versus unconstrained (UC) DCs:} We model the unconstrained DCs in the system by adding their loads to the total ERCOT demand without enforcing the carbon hourly-matching constraint ((\ref{hourly_matching}) - (\ref{flexibility})). 

\begin{figure}[!h]
    \centering
    \includegraphics[width=3.5in]{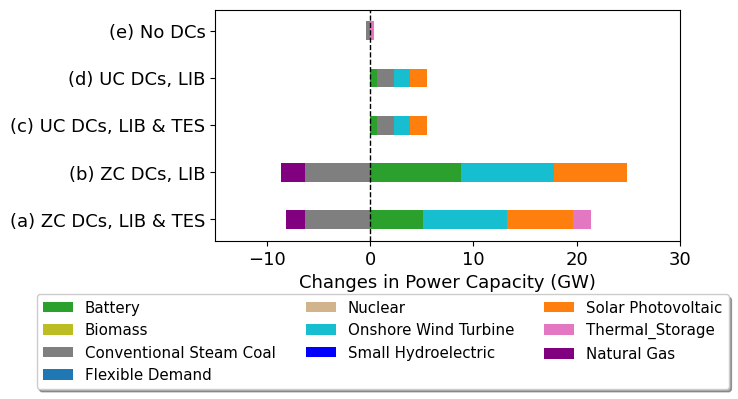}
        \caption{Changes in power capacity (GW) in five scenarios compared to the 2030 ERCOT system without the additional DC loads and coal retrofitting}
    \label{capacity_compare}
\end{figure}

Fig. \ref{capacity_compare} shows changes in power capacity in five scenarios (a) - (e)\footnote{The TES-only scenario is excluded because more than 200GW wind capacity is planned, which is impractical considering the horizon by 2030.} compared to the 2030 ERCOT system without the additional DC loads and coal retrofitting option. First, retrofitting coal into TES is economically viable for the scenario with no DCs (e). It shows that part of the coal capacity is retrofitted for a lower system cost. Second, the retrofitted TES could reduce investments in LIBs, especially under zero-carbon DC operations, as highlighted by comparing scenarios with and without the retrofitted TES (scenarios (a) versus (b), (c) versus (d)). Third, zero-carbon DC systems require much more renewable capacity investment than unconstrained DCs, but the latter leads to retention of the existing coal capacity.

\begin{figure}[!h]
    \centering
    \includegraphics[width=3.5in]{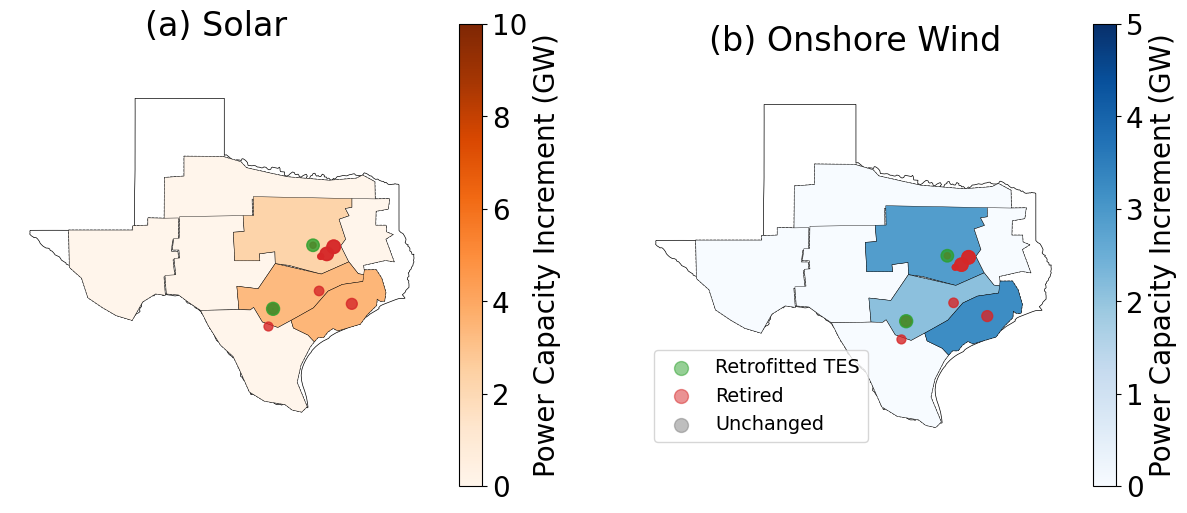}
        \caption{The map of individual retrofitting results in scenario (a) zero-carbon DCs with retrofitted TES, LIB, and co-located renewable generations; DCs are assumed co-located with retrofitted TES}
    \label{Retrofitting_map}
\end{figure}

\begin{figure}[!h]
    \centering
    \includegraphics[width=3.3in]{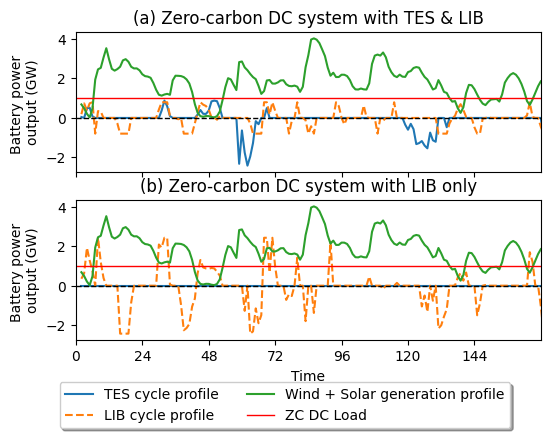}
        \caption{Energy storage cycle and renewable generation profiles of two scenarios (a) TES \& LIB zero-carbon DC systems (b) LIB zero-carbon DC systems; Battery discharging power is positive, and charging power is negative}
    \label{LIB_cycle}
\end{figure}

Fig. \ref{Retrofitting_map} presents the individual retrofitting result of scenario (a) zero-carbon DCs with TES \& LIB. Two of the 12 coal power plants are retrofitted, and the rest are retired. They are coal power plants 8 and 11 in the North and South central zones (Fig. \ref{retrofitted_coal_investment_cost}) with a relatively longer lifetime.  Fig. \ref{LIB_cycle} shows how the complementary operations of TES and LIB support DCs. TES could reduce the LIB capacity investment and the frequent LIB deep cycles under zero-carbon operations. The asymmetrical TES gets charged, stores the surplus renewable energy, and releases it when the renewable power output becomes low.

 \begin{figure}[!h]
    \centering
    \includegraphics[width=3.5in]{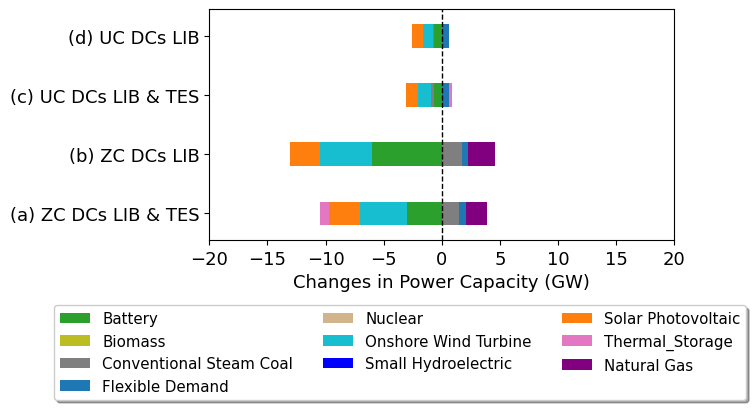}
        \caption{Differences in power capacity (GW) between scenarios with and without flexible loads in DCs}
    \label{changes_in_powercapacity_flex}
\end{figure}

% \begin{figure}[!h]
%    \centering
%    \includegraphics[width=3.5in]{image/changes_in_energy_barchart_flexible.png}
%        \caption{Differences in electricity generations (GWh) between scenarios with and without flexible loads in DCs}
%    \label{changes_in_energy_flex}
%\end{figure}

2) \textit{The impacts of flexible DC operations:} We simulate scenarios in which each DC has $1-\lambda_{z} = 20\%$ of flexible loads (i.e., 200 MW flexible load). Fig. \ref{changes_in_powercapacity_flex} show differences in power capacity of scenarios with and without flexible DC loads. The flexibility of DCs will reduce renewable and energy storage investments. However, this will leave coal capacity unchanged for zero-carbon DC systems (i.e., scenarios (a) and (b)). 

%We compare five system design scenarios with the baseline 2030 system that has no new DCs (represented as $\times 100\%$ baseline) for evaluations. 

% \begin{figure}[!h]
%    \centering
%    \includegraphics[width=3.5in]{image/renewable_curtailment.png}
%        \caption{Percentage changes of five scenarios in renewable curtailment (\%) relative to the baseline case}
%    \label{renewable_curtailment}
%\end{figure}
% 
% 
% \begin{figure}[!h]
%    \centering
%    \includegraphics[width=3.5in]{image/carbon_emission.png}
%        \caption{Percentage changes of five scenarios in carbon emission (\%) relative to the baseline case}
%    \label{carbon_emission}
%\end{figure}

% \begin{figure}[!h]
%    \centering
%    \includegraphics[width=3.5in]{image/curtailment_emission.png}
%        \caption{Percentage changes (\%) of five scenarios in (a) renewable curtailment and (b) carbon emissions compared to the baseline case}
%    \label{renewable_curtailment}
%\end{figure}

 \begin{figure}[!h]
    \centering
    \includegraphics[width=3.5in]{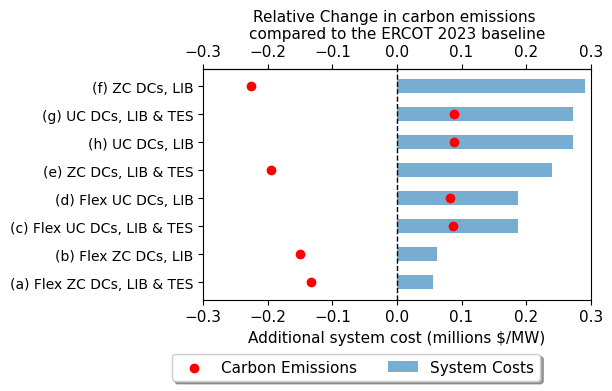}
        \caption{Additional system costs for the per-MW DC load relative to the cost of 2030 ERCOT system that has coal retrofitting but no additional DC loads}
    \label{system_cost}
\end{figure}

Fig. \ref{system_cost} summarizes the system investment cost for the per-MW additional DC load, ranging from \$ 0.055 to \$ 0.291 million. The retrofitting options reduce system cost significantly, and under several scenarios (i.e., (a), (b), and (e)), zero-carbon DC systems are cheaper than unconstrained ones due to the optimal location and configuration. Zero-carbon DCs can achieve 13 - 23\% of carbon reduction.

\section{Conclusions}

We propose a retrofitting strategy that transforms coal plants into TES to support co-located DCs using renewable resources. Complementary operations of retrofitted TES and LIB could reduce system costs and facilitate coal phase-out. Using a capacity expansion model, we show that zero-carbon DCs reusing coal plants could be cheaper than unconstrained DCs with an additional 13 - 23\% carbon reduction. 
\section{Acknowledgment}

The authors would like to thank the funding support from IHI Corporation, Japan.

\bibliographystyle{IEEEtran}
\bibliography{bibtex/IEEEconference}

\begin{thebibliography}{10}
\providecommand{\url}[1]{#1}
\csname url@samestyle\endcsname
\providecommand{\newblock}{\relax}
\providecommand{\bibinfo}[2]{#2}
\providecommand{\BIBentrySTDinterwordspacing}{\spaceskip=0pt\relax}
\providecommand{\BIBentryALTinterwordstretchfactor}{4}
\providecommand{\BIBentryALTinterwordspacing}{\spaceskip=\fontdimen2\font plus
\BIBentryALTinterwordstretchfactor\fontdimen3\font minus
  \fontdimen4\font\relax}
\providecommand{\BIBforeignlanguage}[2]{{%
\expandafter\ifx\csname l@#1\endcsname\relax
\typeout{** WARNING: IEEEtran.bst: No hyphenation pattern has been}%
\typeout{** loaded for the language `#1'. Using the pattern for}%
\typeout{** the default language instead.}%
\else
\language=\csname l@#1\endcsname
\fi
#2}}
\providecommand{\BIBdecl}{\relax}
\BIBdecl

\bibitem{eia_capacity_2023}
\BIBentryALTinterwordspacing
{EIA}, ``Capacity {Factors} for {Utility} {Scale} {Generators} {Primarily}
  {Using} {Fossil} {Fuels},'' May 2023. [Online]. Available:
  \url{https://www.eia.gov/electricity/monthly}
\BIBentrySTDinterwordspacing

\bibitem{lawrence_berkeley_national_laboratory_electricity_market_and_policy_queued_2022}
\BIBentryALTinterwordspacing
{Lawrence Berkeley National Laboratory, Electricity market and policy},
  ``Queued {Up}: {Characteristics} of {Power} {Plants} {Seeking} {Transmission}
  {Interconnection},'' 2022. [Online]. Available:
  \url{https://emp.lbl.gov/queues}
\BIBentrySTDinterwordspacing

\bibitem{yong_retrofitting_2022}
Q.~Yong, Y.~Tian, X.~Qian, and X.~Li, ``\BIBforeignlanguage{en}{Retrofitting
  coal-fired power plants for grid energy storage by coupling with thermal
  energy storage},'' \emph{\BIBforeignlanguage{en}{Applied Thermal
  Engineering}}, vol. 215, p. 119048, Oct. 2022.

\bibitem{hansen_investigating_2022}
\BIBentryALTinterwordspacing
J.~Hansen, W.~Jenson, A.~Wrobel, N.~Stauff, K.~Biegel, T.~Kim, R.~Belles, and
  F.~Omitaomu, ``Investigating {Benefits} and {Challenges} of {Converting}
  {Retiring} {Coal} {Plants} into {Nuclear} {Plants},'' Tech. Rep.
  INL/RPT-22-67964-Rev000, 1886660, Sep. 2022. [Online]. Available:
  \url{https://www.osti.gov/servlets/purl/1886660/}
\BIBentrySTDinterwordspacing

\bibitem{andrew_coffman_smith_plan_2023}
\BIBentryALTinterwordspacing
{Andrew Coffman Smith}, ``Plan unveiled to repurpose {New} {York}'s 2 coal
  plant sites for data centers,'' Sep. 2023. [Online]. Available:
  \url{https://www.spglobal.com/marketintelligence/en/news-insights/trending/$6U9JJYtiN_G7Zo3-a3uYTQ2$}
\BIBentrySTDinterwordspacing

\bibitem{sepulveda_design_2021}
\BIBentryALTinterwordspacing
N.~A. Sepulveda, J.~D. Jenkins, A.~Edington, D.~S. Mallapragada, and R.~K.
  Lester, ``\BIBforeignlanguage{en}{The design space for long-duration energy
  storage in decarbonized power systems},''
  \emph{\BIBforeignlanguage{en}{Nature Energy}}, vol.~6, no.~5, pp. 506--516,
  Mar. 2021. [Online]. Available:
  \url{https://www.nature.com/articles/s41560-021-00796-8}
\BIBentrySTDinterwordspacing

\bibitem{albertus_long-duration_2020}
\BIBentryALTinterwordspacing
P.~Albertus, J.~S. Manser, and S.~Litzelman,
  ``\BIBforeignlanguage{en}{Long-{Duration} {Electricity} {Storage}
  {Applications}, {Economics}, and {Technologies}},''
  \emph{\BIBforeignlanguage{en}{Joule}}, vol.~4, no.~1, pp. 21--32, Jan. 2020.
  [Online]. Available:
  \url{https://linkinghub.elsevier.com/retrieve/pii/S2542435119305392}
\BIBentrySTDinterwordspacing

\bibitem{masanet_recalibrating_2020}
\BIBentryALTinterwordspacing
E.~Masanet, A.~Shehabi, N.~Lei, S.~Smith, and J.~Koomey,
  ``\BIBforeignlanguage{en}{Recalibrating global data center energy-use
  estimates},'' \emph{\BIBforeignlanguage{en}{Science}}, vol. 367, no. 6481,
  pp. 984--986, Feb. 2020. [Online]. Available:
  \url{https://www.science.org/doi/10.1126/science.aba3758}
\BIBentrySTDinterwordspacing

\bibitem{jones_how_2018}
\BIBentryALTinterwordspacing
N.~Jones, ``\BIBforeignlanguage{en}{How to stop data centres from gobbling up
  the world’s electricity},'' \emph{\BIBforeignlanguage{en}{Nature}}, vol.
  561, no. 7722, pp. 163--166, Sep. 2018. [Online]. Available:
  \url{https://www.nature.com/articles/d41586-018-06610-y}
\BIBentrySTDinterwordspacing

\bibitem{mytton_sources_2022}
\BIBentryALTinterwordspacing
D.~Mytton and M.~Ashtine, ``\BIBforeignlanguage{en}{Sources of data center
  energy estimates: {A} comprehensive review},''
  \emph{\BIBforeignlanguage{en}{Joule}}, vol.~6, no.~9, pp. 2032--2056, Sep.
  2022. [Online]. Available:
  \url{https://linkinghub.elsevier.com/retrieve/pii/S2542435122003580}
\BIBentrySTDinterwordspacing

\bibitem{anna_cybulsky_producing_2023}
\BIBentryALTinterwordspacing
{Michael Giovanniello}, {Anna Cybulsky}, {Tim Schittekatte}, and {Dharik S.
  Mallapragada}, ``Producing hydrogen from electricity: {How} modeling
  additionality drives the emissions impact of time-matching requirements,''
  Jan. 2023. [Online]. Available:
  \url{https://energy.mit.edu/wp-content/uploads/2023/04/MITEI-WP-2023-02.pdf}
\BIBentrySTDinterwordspacing

\bibitem{google_data_center_247_nodate}
\BIBentryALTinterwordspacing
{Google Data Center}, ``24/7 {Carbon}-{Free} {Energy} by 2030.'' [Online].
  Available: \url{https://sustainability.google/progress/energy/}
\BIBentrySTDinterwordspacing

\bibitem{zheng_mitigating_2020}
\BIBentryALTinterwordspacing
J.~Zheng, A.~A. Chien, and S.~Suh, ``\BIBforeignlanguage{en}{Mitigating
  {Curtailment} and {Carbon} {Emissions} through {Load} {Migration} between
  {Data} {Centers}},'' \emph{\BIBforeignlanguage{en}{Joule}}, vol.~4, no.~10,
  pp. 2208--2222, Oct. 2020. [Online]. Available:
  \url{https://linkinghub.elsevier.com/retrieve/pii/S2542435120303470}
\BIBentrySTDinterwordspacing

\bibitem{lin_adapting_2023}
\BIBentryALTinterwordspacing
L.~Lin and A.~A. Chien, ``\BIBforeignlanguage{en}{Adapting {Datacenter}
  {Capacity} for {Greener} {Datacenters} and {Grid}},'' in
  \emph{\BIBforeignlanguage{en}{Proceedings of the 14th {ACM} {International}
  {Conference} on {Future} {Energy} {Systems}}}.\hskip 1em plus 0.5em minus
  0.4em\relax Orlando FL USA: ACM, Jun. 2023, pp. 200--213. [Online].
  Available: \url{https://dl.acm.org/doi/10.1145/3575813.3595197}
\BIBentrySTDinterwordspacing

\bibitem{ahmed_review_2021}
\BIBentryALTinterwordspacing
K.~M.~U. Ahmed, M.~H.~J. Bollen, and M.~Alvarez, ``A {Review} of {Data}
  {Centers} {Energy} {Consumption} and {Reliability} {Modeling},'' \emph{IEEE
  Access}, vol.~9, pp. 152\,536--152\,563, 2021. [Online]. Available:
  \url{https://ieeexplore.ieee.org/document/9599719/}
\BIBentrySTDinterwordspacing

\bibitem{gnibga_renewable_2023}
\BIBentryALTinterwordspacing
W.~E. Gnibga, A.~Blavette, and A.-C. Orgerie, ``Renewable {Energy} in {Data}
  {Centers}: the {Dilemma} of {Electrical} {Grid} {Dependency} and {Autonomy}
  {Costs},'' \emph{IEEE Transactions on Sustainable Computing}, pp. 1--13,
  2023. [Online]. Available:
  \url{https://ieeexplore.ieee.org/document/10227588/}
\BIBentrySTDinterwordspacing

\bibitem{noauthor_genx_nodate}
\BIBentryALTinterwordspacing
Y.~Ding, ``{GenX} retrofitting module.'' [Online]. Available:
  \url{https://github.com/GenXProject/GenX/tree/GenX\_retrofit\_MIT}
\BIBentrySTDinterwordspacing

\bibitem{he_optimal_2016}
\BIBentryALTinterwordspacing
G.~He, Q.~Chen, C.~Kang, and Q.~Xia, ``Optimal {Offering} {Strategy} for
  {Concentrating} {Solar} {Power} {Plants} in {Joint} {Energy}, {Reserve} and
  {Regulation} {Markets},'' \emph{IEEE Transactions on Sustainable Energy},
  vol.~7, no.~3, pp. 1245--1254, Jul. 2016. [Online]. Available:
  \url{https://ieeexplore.ieee.org/document/7437454/}
\BIBentrySTDinterwordspacing

\bibitem{staffell_using_2016}
\BIBentryALTinterwordspacing
I.~Staffell and S.~Pfenninger, ``\BIBforeignlanguage{en}{Using bias-corrected
  reanalysis to simulate current and future wind power output},''
  \emph{\BIBforeignlanguage{en}{Energy}}, vol. 114, pp. 1224--1239, Nov. 2016.
  [Online]. Available:
  \url{https://linkinghub.elsevier.com/retrieve/pii/S0360544216311811}
\BIBentrySTDinterwordspacing

\bibitem{us_energy_information_administration_coal-fired_nodate}
\BIBentryALTinterwordspacing
{US Energy Information Administration}, ``Coal-fired electric power plants.''
  [Online]. Available: \url{https://www.eia.gov/coal/data.php\#prices}
\BIBentrySTDinterwordspacing

\bibitem{baxtel_texas_nodate}
\BIBentryALTinterwordspacing
{Baxtel}, ``Texas {Data} {Center} {Map}.'' [Online]. Available:
  \url{https://baxtel.com/data-center/texas}
\BIBentrySTDinterwordspacing

\bibitem{joshua_d_rhodes_impacts_2021}
\BIBentryALTinterwordspacing
{Joshua D. Rhodes}, {Thomas Deetjen}, and {Caitlin Smith}, ``Impacts of
  {Large}, {Flexible} {Data} {Center} {Operations} on the {Future} of
  {ERCOT},'' Jun. 2021. [Online]. Available:
  \url{https://www.ideasmiths.net/wp-content/uploads/2022/02/LANCIUM\textunderscore
  IS\textunderscore ERCOT\textunderscore flexDC\textunderscore
  FINAL\textunderscore 2021.pdf}
\BIBentrySTDinterwordspacing

\bibitem{eia_capital_2020}
\BIBentryALTinterwordspacing
{EIA}, ``Capital {Cost} and {Performance} {Characteristic} {Estimates} for
  {Utility} {Scale} {Electric} {Power} {Generating} {Technologies},'' Feb.
  2020. [Online]. Available:
  \url{https://www.eia.gov/analysis/studies/powerplants/capitalcost/pdf/capital\textunderscore
  cost\textunderscore \\ AEO2020.pdf}
\BIBentrySTDinterwordspacing

\bibitem{ding_supplement_nodate}
\BIBentryALTinterwordspacing
``Supplement material.'' [Online]. Available:
  \url{https://github.com/yifueve/coalrepurpose}
\BIBentrySTDinterwordspacing

\end{thebibliography}

% that's and p all folks
\end{document}